\documentclass[12pt,journal,twocolumn,comsoc]{IEEEtran}
\usepackage[T1]{fontenc}

\usepackage{cite}

\ifCLASSINFOpdf
  \usepackage[pdftex]{graphicx}
  \graphicspath{{../figures/}}
  \DeclareGraphicsExtensions{.pdf,.jpeg,.png}
\else
  \usepackage[dvips]{graphicx}
  \graphicspath{{../figures/}}
  \DeclareGraphicsExtensions{.eps}
\fi
\usepackage{caption}
\usepackage{subcaption}
\usepackage{afterpage}

\usepackage{textcomp}

\usepackage{amsmath,amssymb,amsfonts}
\usepackage{mathtools}
\mathtoolsset{showonlyrefs}

\usepackage[cmintegrals]{newtxmath}
\usepackage{array}
\usepackage{makecell}
\newtheorem{theorem}{Theorem}
\newtheorem{lemma}{Lemma}

\newtheorem{remark}{Remark}

\begin{document}
%
\title{Weakly Private Information Retrieval \\ Under R\'{e}nyi Divergence}
%
%
%
\author{{Jun-Woo~Tak,~\IEEEmembership{Student Member,~IEEE,}
        Sang-Hyo~Kim,~\IEEEmembership{Member,~IEEE,}
        Yongjune~Kim,~\IEEEmembership{Member,~IEEE,}
        and~Jong-Seon~No,~\IEEEmembership{Fellow,~IEEE}}
\thanks{Jun-Woo Tak and Jong-Seon No are with the Department of Electrical and Computer Engineering, Seoul National University, Seoul 08826, South Korea}
\thanks{Sang-Hyo Kim is with the College of Information and Communication Engineering, Sungkyunkwan University, Suwon 16419, South Korea}%
\thanks{Yongjune Kim is with the Department of Information and Communication Engineering, DGIST, Daegu 42988, South Korea}%
\thanks{(Corresponding authors: Sang-Hyo Kim and Yongjune Kim)}}%
\maketitle
\thispagestyle{empty} 

\begin{abstract}
Private information retrieval (PIR) is a protocol that guarantees the privacy of a user who is in communication with databases. The user wants to download one of the messages stored in the databases while hiding the identity of the desired message. Recently, the benefits that can be obtained by weakening the privacy requirement have been studied, but the definition of weak privacy needs to be elaborated upon. In this paper, we attempt to quantify the
weak privacy (i.e., information leakage) in PIR problems by using the R\'{e}nyi divergence that generalizes the Kullback-Leibler divergence. By introducing R\'{e}nyi divergence into the existing PIR problem, the tradeoff relationship between privacy (information leakage) and PIR performance (download cost) is characterized via convex optimization. Furthermore, we propose an alternative PIR scheme with smaller message sizes than the Tian-Sun-Chen (TSC) scheme. The proposed scheme cannot achieve the PIR capacity of perfect privacy since the message size of the TSC scheme is the minimum to achieve the PIR capacity. However, we show that the proposed scheme can be better than the TSC scheme in the \emph{weakly} PIR setting, especially under a low download cost regime. 
\end{abstract}


%
\IEEEpeerreviewmaketitle

\section{Introduction}
%
%
%
%
\IEEEPARstart{P}{rivate} information retrieval (PIR) is a communication protocol presented in seminal work \cite{PIR_ori2} and is devised for the secure use of information stored in databases. In the PIR problem, a user wants to download one of the messages from multiple databases. A user sends queries to the databases and needs to hide what message is being downloaded from databases. Recently, Sun and Jafar \cite{PIR_cap} characterized the information-theoretic capacity of PIR and presented a capacity-achieving scheme. Since then, many studies on PIR have been actively conducted for diverse assumptions and environments \cite{PIR_coded, TPIR, PIR_cache1, PIR_cache2, PIR_MDS, PIR_Byz, PIR_Lsize, PIR_symm, TSC, semanticPIR}.

Among the several derivatives of the classical PIR, recent studies \cite{LPIR, ALPIR, WPIR, WPIR_multi, WPIR_single_C, WPIR_maxL} investigated information leakage in PIR problems in order to achieve lower communication costs. We refer to such PIR scenarios as weakly PIR (WPIR). In \cite{LPIR} and \cite{ALPIR}, by the name of leaky PIR, information leakage was defined similarly to differential privacy. In \cite{WPIR,WPIR_multi,WPIR_single_C}, they proposed the mutual information or the maximal leakage metric as an information leakage measure. Also in \cite{WPIR_maxL} independently, a weakly PIR problem is formulated with the maximal leakage metric by taking into account the Tian-Sun-Chen (TSC) scheme \cite{TSC}. In such studies with information leakage, privacy is relaxed in exchange for the improved download cost. 

In practical operations, we would select the desired privacy level and minimize the download cost. Therefore, one can think of finding an appropriate compromise between information leakage and the download cost. However, the tradeoff between privacy and the performance shown in recent studies does not, by itself, represent the capacity of WPIR. In \cite{ALPIR,WPIR_multi}, the authors exhibited an upper bound and a lower bound on the capacity, but there exists a gap between them. In \cite{WPIR_single_C}, the capacity of WPIR is characterized only for the single-server case.

In this paper, we propose to use the R\'{e}nyi divergence between query distributions as an information leakage measure in the WPIR problem. Then, we formulate the WPIR problem as a convex optimization problem. The problem is reduced to equivalent R\'{e}nyi entropy maximization with arbitrary order. Interestingly, the optimal solution to minimize the R\'{e}nyi divergence (i.e., maximize the R\'{e}nyi entropy) turns out to be the same as the optimum solution to minimize the maximum leakage if we adopt the TSC scheme as in \cite{WPIR_maxL}.


The information-theoretic capacity of PIR was established in \cite{PIR_cap}. However, the capacity of \emph{weakly} PIR (or the fundamental tradeoff between the information leakage and the download cost) is still obscure. We provide evidence that the capacity-achieving schemes of PIR with perfect privacy would not be optimal in the WPIR problem. 

We propose a variant of the capacity-achieving TSC scheme. The proposed scheme has a smaller message size than that of the TSC scheme. Since the message size of the TSC scheme is the minimum value to achieve the capacity, the proposed scheme cannot achieve the PIR capacity of perfect privacy. However, in WPIR, information leakage by the proposed scheme can be lower than that of the TSC scheme's information leakage. The benefit of the proposed scheme is distinct for the lower download cost. 

The rest of the paper is organized as follows. In Section II, preliminaries including a brief review of the PIR problem and its relevant studies on information leakage are given. In Section III, information leakage in PIR is measured by the R\'{e}nyi divergence, and the optimal tradeoff between the information leakage and the download cost is characterized. We present the alternative PIR scheme which achieves lower information leakage than the existing PIR scheme in the low download cost region in Section IV. Lastly, we conclude the paper in Section V.

\section{Preliminaries}
Throughout the paper, we use the following notations. For positive integers $i$ and $j$, where $i\leq j$, we denote $[i:j]=\{i,i+1,\cdots,j\}$ and $A_{[i:j]}=\{A_i,A_{i+1},\cdots,A_j\}$. For a set $A$, we let $A\backslash \{b\}$ denote the set $A$ excluding the element $b$.

\subsection{Classical PIR}

\begin{figure}[h] 
\centering
\includegraphics[scale=0.5]{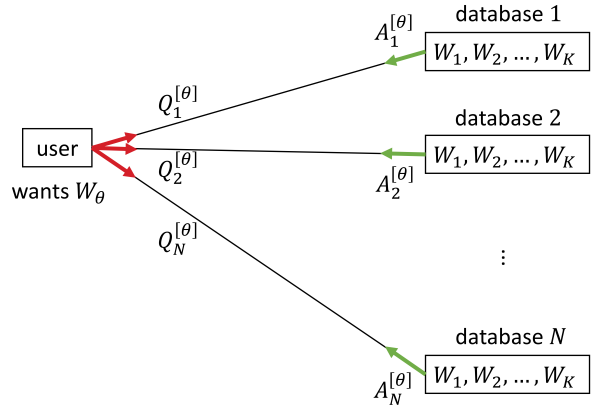}
\caption{Classical PIR scenario with $N$ databases and $K$ messages.}
\label{PIR_fig}
\end{figure}

In the classical PIR scenario, we consider $K$ messages stored identically in $N$ non-colluding databases, as shown in Fig. \ref{PIR_fig}. The messages are denoted by $W_1, W_2, \cdots, W_K$ respectively and equally sized as $H(W_k)=L, k\in[1:K]$, where $H(\cdot)$ denotes the entropy. The purpose of the protocol is for a user to be able to retrieve one of the messages while hiding the wanted message index $\theta$ from all databases. The user sends a query $Q^{[\theta]}_n$ to the database $n\in [1:N]$ and upon receiving the query, the database gives a corresponding answer $A^{[\theta]}_n$ back to the user. The download cost $D_c$ for the user to download the answers in the classical PIR is given as
\begin{equation}
    D_c=\sum^N_{n=1}H(A^{[\theta]}_n).
\end{equation}
The following two conditions should be satisfied in the classical PIR problem. First, the correctness condition guarantees the user's decodability of the message, which is represented as
\begin{equation}
H(W_\theta \mid Q^{[\theta]}_{[1:N]}, A^{[\theta]}_{[1:N]})=0, ~\theta\in [1:K].
\label{correctness}
\end{equation}
Second, the privacy condition that hides the message index from the databases, is often represented by the mutual information between the query and the message index as 
\begin{equation}
I(\theta ; Q^{[\theta]}_n)=0, \theta \in [1:K], ~n\in [1:N].
\label{privacy}
\end{equation}
The performance of the PIR is measured by the rate which is defined as the desired message length divided by the total download cost or $L/D_c$. The capacity of PIR denoted by $C$ is defined as the maximum possible achievable rate. From \cite{PIR_cap}, the capacity of the classical PIR setting is proven to be
\begin{equation}\label{C_PIR}
    C=\left( 1+\frac{1}{N}+\frac{1}{N^2}+\cdots+\frac{1}{N^{K-1}} \right)^{-1}.
\end{equation}

\subsection{Information Leakage in PIR}
Before we discuss the WPIR problems, it is worth mentioning the work of a PIR scheme with perfect privacy in \cite{TSC}, which reduces the conventional message length $L=N^K$ \cite{PIR_cap} to its minimum value $N-1$. In that study, a PIR scheme with probabilistic queries was proposed rather than the conventional deterministic PIR scheme. By using a random key, a user decides a query from a set of choices probabilistically, assuming that there are $M$ options for a user to retrieve the desired message, and any selected option provides the user the desired message equivalently. When the $m$-th option is used to retrieve the message $W_\theta$, we denote the query received at database $n$ and the answer to that query by $Q^{[\theta]}_{n}(m)$ and $A^{[\theta]}_{n}(m)$, respectively. Then the correctness condition becomes
\begin{equation}
\begin{aligned}
    H(W_\theta \mid Q^{[\theta]}_{[1:N]}(m), A^{[\theta]}_{[1:N]}(m))=0, \\
    ~m\in [1:M], ~\theta\in [1:K].
\end{aligned}
\end{equation}
A probabilistic query structure of PIR with options is shown in Table \ref{table1}. $M$ query sets are selected with the probability distribution $P=(p_1, p_2, \cdots, p_M)$ where $p_i \ge 0$ is the probability of the $i$-th query set and $d_i$ is its download cost.

\begin{table}[t] \caption{A probabilistic PIR query structure to retrieve $W_\theta$} \label{table1}
\begin{center}
\resizebox{250pt}{!}{
\begin{tabular}{|c|c|c|c|c|c|}
\hline
 Option & DB $1$ & $\cdots$ & DB $N$ & Probability & \makecell{Download \\cost} \\
\hline
$1$ & $Q^{[\theta]}_1(1)$ & $\cdots$ & $Q^{[\theta]}_N(1)$ & $p_1$ & $d_1$ \\
\hline
$2$ & $Q^{[\theta]}_1(2)$ & $\cdots$ & $Q^{[\theta]}_N(2)$ & $p_2$ & $d_2$ \\
\hline
\vdots & \vdots & & \vdots & \vdots & \vdots \\
\hline
$M$ & $Q^{[\theta]}_1(M)$ & $\cdots$ & $Q^{[\theta]}_N(M)$ & $p_M$ & $d_M$ \\
\hline
\end{tabular}
}
\end{center}
\end{table}

In \cite{TSC}, the uniform distribution for $M$ options is used to achieve the PIR capacity with perfect privacy. However, in the studies on WPIR, a non-uniform distribution is allowed to introduce information leakage.
In the study called leaky PIR \cite{LPIR}, the authors defined the information leakage which was referred to as $\epsilon$-privacy given as
\begin{equation}
    \frac{Pr(Q^{[k_1]}_{n}=q)}{Pr(Q^{[k_2]}_{n}=q)} \leq e^\epsilon, ~k_1,k_2\in[1:K], ~n\in[1:N],
\end{equation}
where non-negative $\epsilon$ bounds the amount of information leakage allowance and $q$ represents a possible realization of the query sent. Weakly PIR \cite{WPIR} defined information leakage in a more information-theoretic manner as nonzero mutual information in privacy condition as
\begin{equation}
    I(\theta ; Q^{[\theta]}_n) \leq \rho, ~\theta \in [1:K], ~n\in [1:N],
\end{equation}
where $\rho$ indicates the amount of information leakage allowance. Also in \cite{WPIR_maxL}, the maximal leakage metric is proposed as a measure of information leakage defined as
\begin{equation}
    \mathcal{L}(\theta\to Q^{[\theta]}_{n})=\log\sum_{q\in \mathcal{Q}}\max_{k\in[1:K]}Pr_{Q^{[k]}_n}(q),
\end{equation}
where $\mathcal{Q}$ is the set of possible queries.

The download cost is defined as the summation of the answer sizes across the databases. The download cost $d_m$ for the $m$-th option is
\begin{align}
    d_m = \sum_{n=1}^{N}{H(A_n^{[\theta]}(m))}.
\end{align}
The performance of this probabilistic PIR structure is measured by the expectation of the download cost normalized by the message length $L$ given as
\begin{align}\label{expected_D}
    D = \frac{1}{L}\sum_{m=1}^{M}{p_m d_m}.
\end{align}
Since $D$ is the amount of download needed per retrieval of unit message length, the achievable rate is its reciprocal.

Essentially, without information leakage, the upper bound of the achievable rate $1/D$ must match the capacity $C$ in \eqref{C_PIR}. However, if the privacy constraint is relaxed, we can reduce the download cost, and thereby achieve a gain in rate. We can consider two extreme cases: perfect privacy and minimal download cost. For the PIR with perfect privacy, the user should download $1/C$ per unit message length. If privacy is not concerned, the user directly downloads the desired message only and the download cost per unit message length is $1$. Therefore, $D$ has the following range depending on the amount of information leakage:
\begin{align}
    D \in \left [1, \frac{1}{C} \right]. \label{D_range}
\end{align}

\subsection{Probabilistic PIR Scheme} \label{subsec3}

\begin{table*}[t] \caption{The probabilistic query structure of symmetric TSC scheme with $N$ databases and $K$ messages when retrieving $W_1$ \cite{TSC}, \cite{semanticPIR}} \label{table_NK}
\begin{center}
\centering 
\begin{tabular}{|c|c|c|c|c|c|c|}
\hline
Option & Database $1$ & Database $2$ & $\cdots$ & Database $N$ & Probability & \thead{Download \\cost} \\
\hline
1 & $W_1(1)$ & $W_1(2)$ & $\cdots$ & $\emptyset$ & $p_1$ & $N-1$ \\
\hline
2 & $\emptyset$ & $W_1(1)$ & $\cdots$ & $W_1(N-1)$ & $p_2$ & $N-1$ \\
\hline
3 & $W_1(N-1)$ & $\emptyset$ & $\cdots$ & $W_1(N-2)$ & $p_3$ & $N-1$ \\
\hline
\vdots & \vdots & \vdots & \vdots & \vdots & \vdots & \vdots \\
\hline
$N$ & $W_1(2)$ & $W_1(3)$ & $\cdots$ & $W_1(1)$ & $p_N$ & $N-1$ \\
\hline
$N+1$ & $W_2(1)$ & $W_1(1)+W_2(1)$ & $\cdots$ & $W_1(N-1)+W_2(1)$ & $p_{N+1}$ & $N$ \\
\hline
$N+2$ & $W_1(N-1)+W_2(1)$ & $W_2(1)$ & $\cdots$ & $W_1(N-2)+W_2(1)$ & $p_{N+2}$ & $N$ \\
\hline
\vdots & \vdots & \vdots & \vdots & \vdots & \vdots & \vdots \\
\hline
$N^K-1$ & $W_1(2)+\displaystyle\sum_{i\in[2:K]}{W_i(N-1)}$  
        & $W_1(3)+\displaystyle\sum_{i\in[2:K]}{W_i(N-1)}$ 
        & $\cdots$ 
        & $W_1(1)+\displaystyle\sum_{i\in[2:K]}{W_i(N-1)}$  
        & $p_{N^K-1}$
        & $N$ \\
\hline
$N^K$ & $W_1(1)+\displaystyle\sum_{i\in[2:K]}{W_i(N-1)}$ 
      & $W_1(2)+\displaystyle\sum_{i\in[2:K]}{W_i(N-1)}$ 
      & $\cdots$ 
      & $\displaystyle\sum_{i\in[2:K]}{W_i(N-1)}$ 
      & $p_{N^K}$ 
      & $N$ \\

\hline
\end{tabular} 
\end{center}
\end{table*}

Throughout this paper, the probabilistic query structure of the TSC scheme proposed in \cite{TSC} is adopted with database symmetry, which corresponds to Scheme 2 in \cite{semanticPIR}. Therefore, we will refer to this scheme as the symmetric TSC scheme. Unlike Scheme 2 in \cite{semanticPIR}, no difference in semantics in messages is assumed in the considered symmetric TSC scheme, and the message length $L$ is fixed to its minimum value $N-1$. Let $W_\theta$ be the desired message with $N-1$ symbols given as
\begin{equation*}
    W_\theta=[W_\theta(1), W_\theta(2), \cdots, W_\theta(N-1)], ~\theta\in[1:K]. 
\end{equation*}
The building process of the probabilistic query structure of PIR is explained as follows \cite{semanticPIR} and Table \ref{table_NK} is given for the retrieval of $W_1$.

\begin{itemize}
    \item (Step~1) Use first $N-1$ databases to download the desired message symbols
    \begin{equation*}
        W_\theta(1), W_\theta(2), \cdots, W_\theta(N-1),
    \end{equation*}
    respectively. Enumerate its cyclic shifts across databases. This step builds first $N$ query options.
    
    \item (Step~2) Download $W_i(1)$ from the first database, where $i\in[1:K]\backslash\{\theta\}$. Use the other $N-1$ databases to download desired message symbols added to $W_i(1)$, that is,
    \begin{align*}
        W_\theta(1)&+W_i(1), \\
        W_\theta(2)&+W_i(1), \\
                   &\cdots, \\
        W_\theta(N-1)&+W_i(1),
    \end{align*}
    respectively. Enumerate its cyclic shifts across databases. Up to the cyclic shifts, we have $N$ query options. Repeat for other symbols in $W_i$, which are $W_i(2), \cdots, W_i(N-1)$. Repeat for each $i\in[1:K]\backslash\{\theta\}$. This step builds $N(N-1)\binom{K-1}{1}$ query options.
    
    \item (Step~3) Download $W_i(1)+W_j(1)$ from the first database, where $i,j\in[1:K]\backslash\{\theta\}$ and $i\ne j$. Use the other $N-1$ databases to download desired message symbols added to $W_i(1)+W_j(1)$, that is,
    \begin{align*}
        W_\theta(1)&+W_i(1)+W_j(1), \\
        W_\theta(2)&+W_i(1)+W_j(1), \\
                   &\quad\quad\cdots, \\ 
        W_\theta(N-1)&+W_i(1)+W_j(1),
    \end{align*}
    respectively. Enumerate its cyclic shifts across databases. Up to the cyclic shifts, we have $N$ query options. Repeat for other symbols in $W_i$ and $W_j$ which are in total $(N-1)^2$ multiple cases. Repeat for other $i,j$, where $i,j\in[1:K]\backslash\{\theta\}$ and $i\ne j$. This step builds $N(N-1)^2\binom{K-1}{2}$ query options.
    
    \item Repeat the steps until it reaches Step $K$. Step $k$ builds $N(N-1)^{k-1}\binom{K-1}{k-1}$ query options.
\end{itemize}

In the query structure, the queries generated in Step 1 trivially request the desired symbols only. It is obvious that the user can obtain the desired message directly. For the queries generated from Step 2 to Step $K$, the user can subtract the undesired symbol or the sum of the undesired symbols from the other symbols containing the desired symbol in order to recover the desired message.

By adding up the number of query options built from each step, the total number of possible options is calculated as
\begin{equation} \label{numopt}
\begin{aligned}
    N &+ N(N-1)\binom{K-1}{1} + \cdots + N(N-1)^{K-1}\binom{K-1}{K-1}\\
    &= N\sum_{k=0}^{K-1}{(N-1)^{k}\binom{K-1}{k}}\\
    &= N\cdot N^{K-1}\\
    &= N^K.
\end{aligned}
\end{equation}
Therefore, there are $N^K$ options from each of which a user can obtain the desired message.

\section{Main Results}

\subsection{Information Leakage Using R\'{e}nyi Divergence}
Consider the probabilistic PIR query structure of Table \ref{table1}. 
Let the probability distribution $U=(u_1, u_2, \cdots, u_M)$ for $M$ options achieve perfect privacy. In \cite{TSC}, the uniform distribution is used for $U$. If a chosen distribution $P$ is used instead of $U$, we can introduce information leakage and reduce the download cost. We propose to use the R\'{e}nyi divergence of order $\alpha$ from $U$ to $P$ as an information leakage measure denoted by $\rho_\alpha$, i.e.,
\begin{align}\label{D_KL_PU}
    \rho_{\alpha} &= D_{\alpha}(P\parallel U) \\
                  &=\frac{1}{\alpha-1} \log \sum_{m=1}^{M}{\frac{p_m^{\alpha}}{u_m^{\alpha-1}}},
                    \label{D_KL_PU2}
\end{align}
where $0<\alpha<\infty$ and $\alpha \ne 1$. The logarithm is used with base $e$. We denote $\lim_{\alpha\to 1}\rho_\alpha$ by $\rho_1$ which corresponds to the Kullback-Leibler (KL) divergence, that is
\begin{align}\label{D_KL_PU_alpha1}
    \rho_{1} &=\sum_{m=1}^{M}{p_m \log \dfrac{p_m}{u_m}}.
\end{align}
Also, as $\alpha\to\infty$, we have 
\begin{align}
    \rho_{\infty} &=\log \max_m \dfrac{p_m}{u_m}.
\end{align}
Note that if and only if $P=U$, then the information leakage equals zero (i.e., perfect privacy). $\rho_\alpha$ indicates how far the chosen $P$ diverges from the state of no leakage.

The goal of this paper is to find the probability distribution $P$ that minimizes the information leakage measured by R\'{e}nyi divergence for a given target download cost $D$. Then, the optimal tradeoff between the download cost and the information leakage can be characterized by solving the optimization problem. We formulate the following convex optimization problem
\begin{equation}\label{optProb}
\begin{aligned}
{\text{minimize}}~~~~~ &\rho_{\alpha} \\
\text{subject to}~~~~~ &\frac{1}{L}\sum_{m=1}^{M}{p_m d_m} = D, \\
                &\sum_{m=1}^{M}{p_m} = 1.
\end{aligned}
\end{equation}
 The optimality of the tradeoff is obtained from the fact that the convex optimization problem \eqref{optProb} has its global optimum, since it has the convex objective function and affine constraint functions.

\subsection{Optimization of Probability Distribution}
First, we will solve the convex optimization problem in \eqref{optProb} for the symmetric TSC scheme. Note that the scheme has $N^K$ options and the message length of $L=N-1$ as in Table \ref{table_NK}. Drawing from the fact that the scheme achieves perfect privacy by using the uniform probability distribution for $U$, the problem in \eqref{optProb} can be rewritten as
\begin{equation} \label{minTSC}
\begin{aligned}
    \underset{P}{\text{minimize}}~~~~~   &\frac{1}{\alpha-1} \log \sum_{m=1}^{N^K}{{p_m^{\alpha}}} + \log N^K \\
    \text{subject to}~~~~~ &\frac{1}{N-1}\sum_{m=1}^{N^K}{p_m d_m} = D, \\
                    &\sum_{m=1}^{N^K}{p_m} = 1,
\end{aligned}
\end{equation}
for $0<\alpha<\infty$ and $\alpha\ne 1$,

\begin{equation} \label{minTSC_KL}
\begin{aligned}
    \underset{P}{\text{minimize}}~~~~~ &\sum_{m=1}^{N^K}{p_m \log p_m} + \log N^K \\
    \text{subject to}~~~~~ &\frac{1}{N-1}\sum_{m=1}^{N^K}{p_m d_m} = D, \\
                    &\sum_{m=1}^{N^K}{p_m} = 1,
\end{aligned}
\end{equation}
for $\alpha=1$, and

\begin{equation} \label{minTSC_infty}
\begin{aligned}
    \underset{P}{\text{minimize}}~~~~~   &\log\max_m{p_m} + \log N^K \\
    \text{subject to}~~~~~ &\frac{1}{N-1}\sum_{m=1}^{N^K}{p_m d_m} = D, \\
                    &\sum_{m=1}^{N^K}{p_m} = 1,
\end{aligned}
\end{equation}
for $\alpha\to\infty$.

Since the objective functions in \eqref{minTSC}, \eqref{minTSC_KL}, and \eqref{minTSC_infty} are in the form of the summation of the negative R\'{e}nyi entropy and a constant, they are convex over their domain. Therefore, each of the three problems has a unique solution, respectively, and the solutions to the first two problems are found to be the same. The solutions are summarized in the following lemma, whose proof is given in Appendix \ref{app.A} and \ref{app.B}.

\begin{lemma}
The optimal solution to the convex optimization problem in \eqref{optProb} on the symmetric TSC scheme is given as \label{lemma1}
    \begin{align}
        p_m = 
        \begin{cases}
            \frac{1-(N-1)(D-1)}{N}, & m\in[1:N], \\
            \frac{(N-1)(D-1)}{N^K-N}, & m\in [N+1:N^K],
        \end{cases}
    \end{align}
for $0<\alpha<\infty$, and

    \begin{align}
        p_m &= \frac{1-(N-1)(D-1)}{N}, ~m\in[1:N],
    \end{align}
and $p_m$ for $m\in [N+1:N^K]$ can be any distribution satisfying $\sum_{m=N+1}^{N^K}{p_m}=(N-1)(D-1)$ and $p_m\leq \frac{1-(N-1)(D-1)}{N}$, for $\alpha\to\infty$.
\end{lemma}

\begin{remark}
The problems of minimizing information leakage expressed in \eqref{minTSC}, \eqref{minTSC_KL}, and \eqref{minTSC_infty} are equivalent to the maximization of the R\'{e}nyi entropy, including the special cases of the Shannon entropy and the min-entropy. We note that minimizing information leakage can be interpreted as maximizing the uncertainty of query options. Intuitively, $p_1 = ... = p_N$ and $p_{N+1} = ... = p_{N^K}$ will maximize the entropy for a given download constraint. Note that the optimized solution in Lemma \ref{lemma1} maximizes the R\'{e}nyi entropy with any $\alpha$.
Also, the solution to $p_m$ for arbitrary $0<\alpha<\infty$ can be the solution to $p_m$ for $\alpha\to\infty$. However, the converse does not hold.

\end{remark}

\begin{remark}
The optimized probability distribution in Lemma 1 is equivalent to the optimal distribution in \cite[Theorem 1]{WPIR_maxL}. It shows that the solution in Lemma 1 minimizes the R\'{e}nyi divergence of any $\alpha$ as well as the maximal leakage metric for a given download cost constraint. Since we consider the symmetric TSC scheme for which all $N$ cyclic shifts of an option across databases are explicitly included in the query options, the probability of each option is $1/N$ of in \cite{WPIR_maxL}.
\end{remark}

\subsection{Optimal Tradeoff Between Information Leakage and Download Cost}
In this subsection, we discuss the optimal tradeoff that can be achieved from Lemma \ref{lemma1}. The following theorem characterizes the optimal tradeoff between the information leakage measured in the R\'{e}nyi divergence $\rho_{\alpha}$ and the expected normalized download cost $D$ on the symmetric TSC scheme.

\begin{theorem}\label{Theorem1}
    The optimal tradeoff between the information leakage measured by the R\'{e}nyi divergence $\rho_{\alpha}$ of order $\alpha$, where $0<\alpha<\infty$ and $\alpha\ne 1$ and the expected normalized download cost $D$, where $1 \leq D \leq \frac{1}{C}$ on the symmetric TSC scheme is given as
    \begin{equation}
    \begin{aligned} \label{thm1}
        \rho_{\alpha} = & ~\frac{1}{\alpha-1}\log
        \biggr[ N \left \{ \frac{1-(N-1)(D-1)}{N} \right \}^{\alpha} \\
        & + (N^K-N)\left \{\frac{(N-1)(D-1)}{N^K-N}\right \}^{\alpha} \biggr] + \log N^K.
    \end{aligned}
    \end{equation}
    Then the information leakage-download cost pairs establish the optimal tradeoff. In the special cases of $\alpha=1$ and $\alpha\to\infty$, $\rho_1$ and $\rho_\infty$ are given as
    \begin{equation}
    \begin{aligned} \label{thm1_KL}
        \rho_{1} = & ~\left\{1-(N-1)(D-1)\right\}\log\frac{1-(N-1)(D-1)}{N} \\
                    & +(N-1)(D-1)\log\frac{(N-1)(D-1)}{N^K-N} \\
                    & +\log N^K,
    \end{aligned}
    \end{equation}
    and
    \begin{equation}
    \begin{aligned} \label{thm1_infty}
        \rho_{\infty} =& \log\frac{1-(N-1)(D-1)}{N} +\log N^K,
    \end{aligned}
    \end{equation}
    respectively.
\end{theorem}

\begin{IEEEproof}
Proof is given by substituting the optimal solution from Lemma \ref{lemma1} in the objective functions of \eqref{minTSC}, \eqref{minTSC_KL}, and \eqref{minTSC_infty}.
\end{IEEEproof}

\begin{table*}[t] \caption{The probabilistic query structure of the symmetric TSC scheme with $N=3,K=2$ to retrieve $W_1$} \label{table_N3K2}
\begin{center}
    \begin{tabular}[b]{ |c|c|c|c|c|c|c| }
    \hline
    Option & Database $1$ & Database $2$ & Database $3$ & Probability & \thead{Download \\cost} \\
    \hline
    1 & $W_1(1)$ & $W_1(2)$ & $\emptyset$ & $p_1$ & 2 \\
    \hline
    2 & $\emptyset$ & $W_1(1)$ & $W_1(2)$ & $p_2$ & 2 \\
    \hline
    3 & $W_1(2)$ & $\emptyset$ & $W_1(1)$ & $p_3$ & 2 \\
    \hline
    4 & $W_2(1)$ & $W_1(1)+W_2(1)$ & $W_1(2)+W_2(1)$ & $p_4$ & 3 \\
    \hline
    5 & $W_1(2)+W_2(1)$ & $W_2(1)$ & $W_1(1)+W_2(1)$ & $p_5$ & 3 \\
    \hline
    6 & $W_1(1)+W_2(1)$ & $W_1(2)+W_2(1)$ & $W_2(1)$ & $p_6$ & 3 \\
    \hline
    7 & $W_2(2)$ & $W_1(1)+W_2(2)$ & $W_1(2)+W_2(2)$ & $p_7$ & 3 \\
    \hline
    8 & $W_1(2)+W_2(2)$ & $W_2(2)$ & $W_1(1)+W_2(2)$ & $p_8$ & 3 \\
    \hline
    9 & $W_1(1)+W_2(2)$ & $W_1(2)+W_2(2)$ & $W_2(2)$ & $p_9$ & 3 \\
    \hline
    \end{tabular}
\end{center}
\end{table*}

\begin{table*}[t] \caption{The probabilistic query structure of the symmetric TSC scheme with $N=4,K=2$ to retrieve $W_1$} \label{table_N4K2}
\begin{center}
    
\begin{tabular}{ |c|c|c|c|c|c|c| }
    \hline
    Option & Database $1$ & Database $2$ & Database $3$ & Database $4$ & Probability & \thead{Download \\cost} \\
    \hline
    1 & $W_1(1)$ & $W_1(2)$ & $W_1(3)$ & $\emptyset$ & $p_1$ & 3 \\
    \hline
    2 & $\emptyset$ & $W_1(1)$ & $W_1(2)$ & $W_1(3)$ & $p_2$ & 3 \\
    \hline
    3 & $W_1(3)$ & $\emptyset$ & $W_1(1)$ & $W_1(2)$ & $p_3$ & 3 \\
    \hline
    4 & $W_1(2)$ & $W_1(3)$ & $\emptyset$ & $W_1(1)$ & $p_4$ & 3 \\
    \hline
    5 & $W_2(1)$ & $W_1(1)+W_2(1)$ & $W_1(2)+W_2(1)$ & $W_1(3)+W_2(1)$ & $p_5$ & 4 \\
    \hline
    6 & $W_2(2)$ & $W_1(1)+W_2(2)$ & $W_1(2)+W_2(2)$ & $W_1(3)+W_2(2)$ & $p_6$ & 4 \\
    \hline
    7 & $W_2(3)$ & $W_1(1)+W_2(3)$ & $W_1(2)+W_2(3)$ & $W_1(3)+W_2(3)$ & $p_7$ & 4 \\
    \hline
    8 & $W_1(3)+W_2(1)$ & $W_2(1)$ & $W_1(1)+W_2(1)$ & $W_1(2)+W_2(1)$ & $p_8$ & 4 \\
    \hline
    9 & $W_1(3)+W_2(2)$ & $W_2(2)$ & $W_1(1)+W_2(2)$ & $W_1(2)+W_2(2)$ & $p_9$ & 4 \\
    \hline
    10& $W_1(3)+W_2(3)$ & $W_2(3)$ & $W_1(1)+W_2(3)$ & $W_1(2)+W_2(3)$ & $p_{10}$ & 4 \\
    \hline
    11& $W_1(2)+W_2(1)$ & $W_1(3)+W_2(1)$ & $W_2(1)$ & $W_1(1)+W_2(1)$ & $p_{11}$ & 4 \\
    \hline
    12& $W_1(2)+W_2(2)$ & $W_1(3)+W_2(2)$ & $W_2(2)$ & $W_1(1)+W_2(2)$ & $p_{12}$ & 4 \\
    \hline
    13& $W_1(2)+W_2(3)$ & $W_1(3)+W_2(3)$ & $W_2(3)$ & $W_1(1)+W_2(3)$ & $p_{13}$ & 4 \\
    \hline
    14& $W_1(1)+W_2(1)$ & $W_1(2)+W_2(1)$ & $W_1(3)+W_2(1)$ & $W_2(1)$ & $p_{14}$ & 4 \\
    \hline
    15& $W_1(1)+W_2(2)$ & $W_1(2)+W_2(2)$ & $W_1(3)+W_2(2)$ & $W_2(2)$ & $p_{15}$ & 4 \\
    \hline
    16& $W_1(1)+W_2(3)$ & $W_1(2)+W_2(3)$ & $W_1(3)+W_2(3)$ & $W_2(3)$ & $p_{16}$ & 4 \\
    \hline
    \end{tabular}
\end{center}
\end{table*}

\begin{remark} \label{remark3}
We can verify two extreme points of the tradeoff: the perfect privacy case and the minimum download case. For the case of perfect privacy, substituting $D=1/C=1+1/N\cdots+1/N^{K-1} = 1+\frac{1-1/N^{K-1}}{N-1}$ in \eqref{thm1} leads to
    \begin{equation} \label{min_rho}
    \begin{aligned}
        \min & (\rho_{\alpha}) \\
        &= \frac{1}{\alpha-1}\log\biggr[ N \left ( \frac{1}{N^K} \right )^{\alpha} \\
        &\quad + (N^K-N)\left ( \frac{1}{N^K} \right )^{\alpha} \biggr] + \log N^K \\
        &= \frac{1}{\alpha-1}\log\left(\frac{1}{N^K}\right)^{\alpha-1} + \log N^K \\
        &= 0.
    \end{aligned}
    \end{equation}
For the case of the minimum download cost (i.e., $D = 1$), \eqref{thm1} becomes
\begin{equation} \label{max_rho}
    \begin{aligned}
        \max(\rho_{\alpha}) =&~ \frac{1}{\alpha-1}\log
        \biggr[ \left ( \frac{1}{N} \right )^{\alpha-1} \biggr] + \log N^K \\
         =&~ \log\frac{1}{N} + \log N^K \\
         =&~ \log N^{K-1},
    \end{aligned}
    \end{equation}
as the maximum leakage value. Similarly, the same observations can be found from \eqref{thm1_KL} and \eqref{thm1_infty}.
\end{remark}

\begin{figure*}[t]
    \centering
    \begin{subfigure}[t]{0.48\textwidth}
        \centering
        \includegraphics[width=\textwidth]{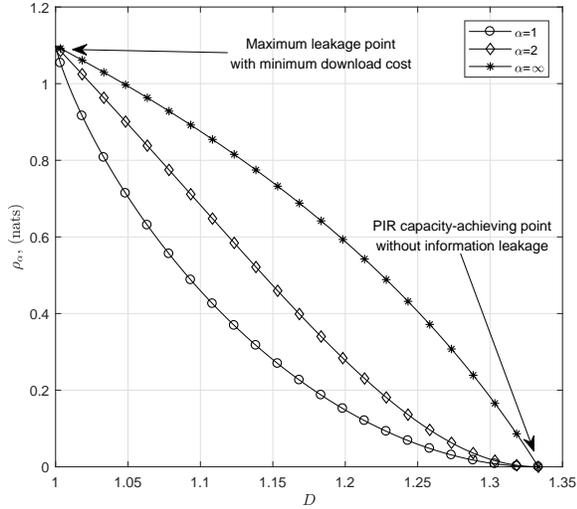}
        \caption{$N=3$, $K=2$}
    \end{subfigure}
    \begin{subfigure}[t]{0.48\textwidth}
        \centering
        \includegraphics[width=\textwidth]{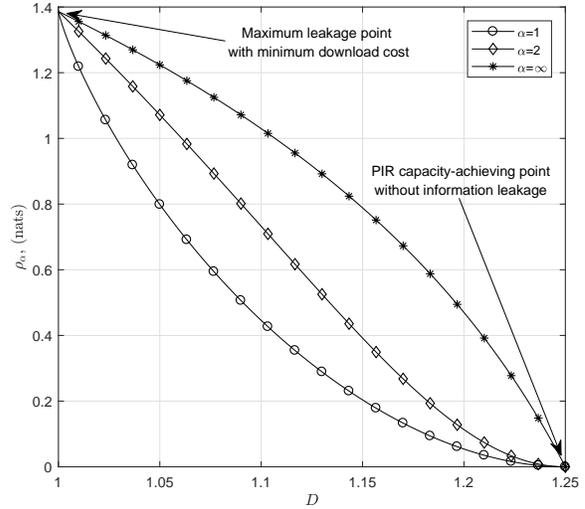}
        \caption{$N=4$, $K=2$}
    \end{subfigure}
\caption{Optimal tradeoffs between information leakage and download cost on the symmetric TSC scheme.}\label{graphN34K2}
\end{figure*}

\begin{figure*}[t]
    \centering
    \begin{subfigure}[t]{0.48\textwidth}
        \centering
        \includegraphics[width=\textwidth]{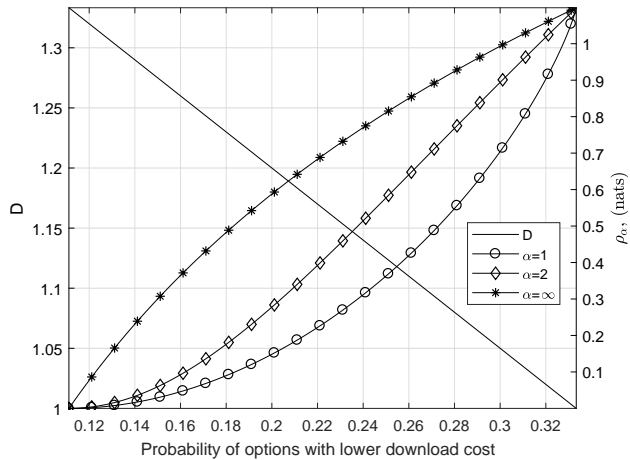}
        \caption{$N=3$, $K=2$}
    \end{subfigure}
    \begin{subfigure}[t]{0.48\textwidth}
        \centering
        \includegraphics[width=\textwidth]{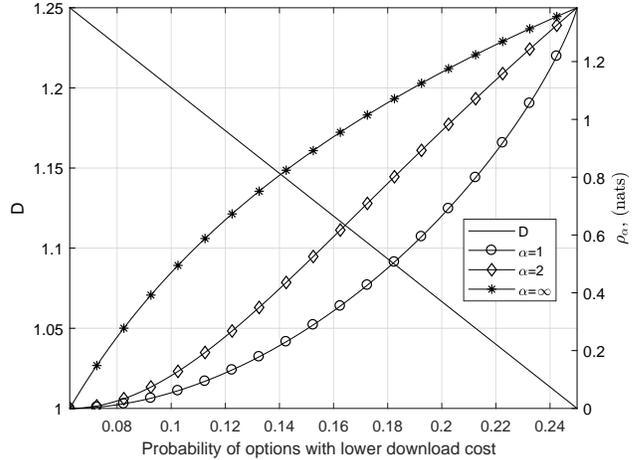}
        \caption{$N=4$, $K=2$}
    \end{subfigure}
\caption{Information leakage and download cost according to the probability of lower download cost on the symmetric TSC scheme.}\label{graphN34K2_X}
\end{figure*}

\subsection{Numerical Analysis}
In this section, we examine our theoretical results by looking at two simple examples of the symmetric TSC scheme. First, Table \ref{table_N3K2} shows the probabilistic query structure for $N=3,K=2$ and $L=2$ to retrieve $W_1$. Note that there are 9 options, and by choosing any of these options, a user can retrieve the required message symbols $W_1(1)$ and $W_1(2)$. By allocating the uniform distribution to $P=(p_1,\cdots,p_9)$, databases will be accessed with all possible queries in an unbiased manner, i.e., resulting in perfect privacy. By allowing information leakage, $p_m\ge p_{m'}$ for $m\in[1:3]$ and $m'\in[4:9]$ from Lemma \ref{lemma1}. Since queries are selected with biased $P$, databases would be able to detect which message the user requests. Fig. \ref{graphN34K2}(a) shows the tradeoff curve between the information leakage $\rho_{\alpha}$ and the download cost $D$. Note that the three typical orders of $\alpha=1,2$, and $\infty$ are presented. The zero leakage point corresponds to the PIR capacity-achieving point without information leakage, where the expected normalized download cost is calculated as
\begin{equation}\label{D_N3K2}
    D=\frac{1}{2} \cdot \frac{1}{9} \cdot (2+2+2+3+3+3+3+3+3)=\frac{4}{3},
\end{equation}
which is the reciprocal of the PIR capacity. The other extreme point is the full information leakage or zero privacy point, where only the first three options with direct downloading are used. With $D=1$, the information leakage calculated in \eqref{max_rho} is given as
\begin{equation}
    \rho_{\alpha}= \log 3^{2-1} \simeq 1.0986~\text{nats}.
\end{equation}

In Table \ref{table_N4K2}, we have the probabilistic query structure for parameters $N=4,K=2$ and $L=3$. There are 16 options with two groups of probabilities of $(p_1,p_2,p_3,p_4)$ and $(p_5,\cdots,p_{16})$. Similar observations of the tradeoff on this example are shown in Fig. \ref{graphN34K2}(b).

We provide a different analysis in Fig. \ref{graphN34K2_X}(a), where information leakage and download cost according to the probability of options with lower download costs are shown for $N=3,K=2$. The probability in the figure corresponds to $p_m$ for $m\in[1:N]$ in Lemma \ref{lemma1}. As the probability of lower download costs increases, we observe that the download cost decreases and information leakage increases. The probability of options ranges from $1/N^K$ to $1/N$, which corresponds to perfect privacy and zero privacy, respectively. Similar analysis on the case of $N=4,K=2$ can be given with Fig. \ref{graphN34K2_X}(b).


\begin{table*}[t] \caption{The probabilistic query structure of the proposed alternative PIR scheme with $N=3,K=2$ to retrieve $W_1$}
\begin{center}
\begin{tabular}{ |c|c|c|c|c|c|c| }
\hline
Option & Database $1$ & Database $2$ & Database $3$ & Probability & \thead{Download \\cost} \\
\hline
1 & $W_1(1)$ & $\emptyset$ & $\emptyset$ & $p_1$ & 1 \\
\hline
2 & $\emptyset$ & $W_1(1)$ & $\emptyset$ & $p_2$ & 1 \\
\hline
3 & $\emptyset$ & $\emptyset$ & $W_1(1)$ & $p_3$ & 1 \\
\hline
4 & $W_2(1)$ & $W_1(1)+W_2(1)$ & $\emptyset$ & $p_4$ & 2 \\
\hline
5 & $\emptyset$ & $W_2(1)$ & $W_1(1)+W_2(1)$ & $p_5$ & 2 \\
\hline
6 & $W_1(1)+W_2(1)$ & $\emptyset$ & $W_2(1)$ & $p_6$ & 2 \\
\hline
\end{tabular}
\end{center}
\vspace{15pt}
\label{table_N3K2_alt}
\end{table*}

\section{Alternative Probabilistic PIR Scheme}
In this section, we propose an alternative PIR scheme. The scheme achieves a better tradeoff between information leakage and download cost within a certain range, although it cannot achieve the perfect privacy PIR capacity. The proposed scheme is obtained by modifying the symmetric TSC scheme so as to obtain a shorter message length, i.e., $L < N - 1$. In \cite{TSC}, it was shown that the optimal message length is $N-1$ in order to achieve the perfect privacy capacity. However, we show that the optimal message length can be lower than $N - 1$ if information leakage is allowed. As we have the example of the symmetric TSC scheme for $N=3,K=2$ from the previous section, we begin with the alternative PIR scheme with the same parameters as in Table \ref{table_N3K2_alt} for comparison. Note that the number of options is reduced from 9 to 6, and the message length $L$ is reduced from 2 to 1. Note that perfect privacy is achieved by the uniform distribution over the 6 query options; databases are accessed with unbiased queries. However, unlike the symmetric TSC scheme, the alternative PIR scheme with the uniform distribution does not achieve the PIR capacity. The corresponding download cost with perfect privacy is given by
\begin{equation}
    D=\frac{1}{6}\cdot (1+1+1+2+2+2)=\frac{3}{2},
\end{equation}
which is greater than the reciprocal of the capacity $3/4$. Interestingly, this alternative PIR scheme can be better than the TSC scheme in the low download cost region. We detail the alternative PIR scheme in the next following subsection.

\subsection{The Proposed Alternative Probabilistic PIR Scheme}
The alternative PIR scheme is based on the equivalent structure of the symmetric TSC scheme, and has shorter message lengths $L$ and smaller option sizes $M$. Let $W_\theta$ be the desired message given as
\begin{equation*}
    W_\theta=[W_\theta(1), W_\theta(2), \cdots, W_\theta(L)], ~\theta\in[1:K],    
\end{equation*}
then the query generation of the alternative PIR scheme is identical to that of the symmetric TSC scheme, but with shorter message lengths $L$, where $1\leq L < N-1$. In Step 1, use the first $L$ databases to download desired message symbols, respectively. Cyclic shifts across databases follow after then. From Step 2 to Step $K$, desired symbols are downloaded after being added with undesired symbols.
By adding up the numbers of options built in each step, the total number of possible options for the proposed alternative PIR scheme is calculated as
\begin{equation} \label{numopt_alt}
\begin{aligned}
    N &+ NL\binom{K-1}{1} + \cdots + NL^{K-1}\binom{K-1}{K-1}\\
    &= N\sum_{k=0}^{K-1}{L^{k}\binom{K-1}{k}}\\
    &= N(L+1)^{K-1}.
\end{aligned}
\end{equation}
Since $L<N-1$, the total number of options in this scheme is lower than that of the symmetric TSC scheme, i.e., $N^K$.

\subsection{Optimization}
Now, as in \eqref{minTSC}, \eqref{minTSC_KL}, and \eqref{minTSC_infty}, we can solve the convex optimization problem in \eqref{optProb} for the alternative PIR scheme with the reduced number of options $N(L+1)^{K-1}$ and message length $L<N-1$. Since the alternative PIR scheme inherits the query generation method of the symmetric TSC scheme, the uniform distribution for $N(L+1)^{K-1}$ options gives perfect privacy. The problem in \eqref{optProb} is then rewritten as
\begin{equation} \label{min_alt}
\begin{aligned}
    \underset{P}{\text{minimize}}~ &\frac{1}{\alpha-1} \log \sum_{m=1}^{N(L+1)^{K-1}}{{p_m^{\alpha}}} + \log N(L+1)^{K-1} \\
    \text{subject to}~ &\frac{1}{L}\sum_{m=1}^{N(L+1)^{K-1}}{p_m d_m} = D, \\
                    &\sum_{m=1}^{N(L+1)^{K-1}}{p_m} = 1,
\end{aligned}
\end{equation}
for $0<\alpha<\infty$ and $\alpha\ne 1$,

\begin{equation} \label{min_alt_KL}
\begin{aligned}
    \underset{P}{\text{minimize}}~~~ &\sum_{m=1}^{N(L+1)^{K-1}}{p_m \log p_m} + \log N(L+1)^{K-1} \\
    \text{subject to}~~~ &\frac{1}{L}\sum_{m=1}^{N(L+1)^{K-1}}{p_m d_m} = D, \\
                    &\sum_{m=1}^{N(L+1)^{K-1}}{p_m} = 1,
\end{aligned}
\end{equation}
for $\alpha=1$, and

\begin{equation} \label{min_alt_infty}
\begin{aligned}
    \underset{P}{\text{minimize}}~~~   &\log\max_m{p_m} + \log N(L+1)^{K-1} \\
    \text{subject to}~~~ &\frac{1}{L}\sum_{m=1}^{N(L+1)^{K-1}}{p_m d_m} = D, \\
                    &\sum_{m=1}^{N(L+1)^{K-1}}{p_m} = 1,
\end{aligned}
\end{equation}
for $\alpha\to\infty$.

The problems in \eqref{min_alt}, \eqref{min_alt_KL}, and \eqref{min_alt_infty} can be solved in the same way as \eqref{minTSC}, \eqref{minTSC_KL}, and \eqref{minTSC_infty} for different download costs
\begin{align}
	d_m =
        \begin{cases}
            L, & m\in[1:N], \\
            L+1,  & m\in[N+1:N(L+1)^{K-1}].
        \end{cases} \label{d_m_alt}
\end{align}
We present the solution to the problems in the following lemma without detailed proof. Note that the solution resembles Lemma \ref{lemma1} considering that the message length term $N-1$ is replaced with $L$, and the number of options $N^K$ is replaced with $N(L+1)^{K-1}$.

\begin{lemma} \label{lemma2}
The optimal solution to the convex optimization problem \eqref{optProb} on the alternative probabilistic PIR scheme with message length $L$ satisfying $1\leq L < N-1$ is given as
    \begin{align}
        p_m =
        \begin{cases}
            \frac{1-L(D-1)}{N}, & m\in[1:N], \\
            \frac{L(D-1)}{N(L+1)^{K-1}-N}, & m\in[N+1:N(L+1)^{K-1}],
        \end{cases}
    \end{align}
for $0<\alpha<\infty$, and
    \begin{align}
        p_m &= \frac{1-L(D-1)}{N}, ~m\in[1:N],
    \end{align}
and $p_m$ for $m\in [N+1:N(L+1)^{K-1}]$ can be any distribution satisfying $\sum_{m=N+1}^{N(L+1)^{K-1}}{p_m}=L(D-1)$ and $p_m\leq \frac{1-L(D-1)}{N}$, for $\alpha\to\infty$.
\end{lemma}

\begin{table*}[t] \caption{The probabilistic query structures of the alternative PIR scheme with $N=4,K=2$ to retrieve $W_1$}  \label{table_N4K2_alt}
    \begin{subtable}[h]{\textwidth}
        \centering
        \caption{Message length $L=2$}
        \begin{tabular}{ |c|c|c|c|c|c|c| }
        \hline
        Option & Database $1$ & Database $2$ & Database $3$ & Database $4$ & Probability & \thead{Download \\cost} \\
        \hline
        1 & $W_1(1)$ & $W_1(2)$ & $\emptyset$ & $\emptyset$ & $p_1$ & 2 \\
        \hline
        2 & $\emptyset$ & $W_1(1)$ & $W_1(2)$ & $\emptyset$ & $p_2$ & 2 \\
        \hline
        3 & $\emptyset$ & $\emptyset$ & $W_1(1)$ & $W_1(2)$ & $p_3$ & 2 \\
        \hline
        4 & $W_1(2)$ & $\emptyset$ & $\emptyset$ & $W_1(1)$ & $p_4$ & 2 \\
        \hline
        5 & $W_2(1)$ & $W_1(1)+W_2(1)$ & $W_1(2)+W_2(1)$ & $\emptyset$ & $p_5$    & 3 \\
        \hline
        6 & $W_2(2)$ & $W_1(1)+W_2(2)$ & $W_1(2)+W_2(2)$ & $\emptyset$ & $p_6$    & 3 \\
        \hline
        7 & $\emptyset$ & $W_2(1)$ & $W_1(1)+W_2(1)$ & $W_1(2)+W_2(1)$ & $p_7$    & 3 \\
        \hline
        8 & $\emptyset$ & $W_2(2)$ & $W_1(1)+W_2(2)$ & $W_1(2)+W_2(2)$ & $p_8$    & 3 \\
        \hline
        9 & $W_1(2)+W_2(1)$ & $\emptyset$ & $W_2(1)$ & $W_1(1)+W_2(1)$ & $p_9$    & 3 \\
        \hline
        10& $W_1(2)+W_2(2)$ & $\emptyset$ & $W_2(2)$ & $W_1(1)+W_2(2)$ & $p_{10}$ & 3 \\
        \hline
        11& $W_1(1)+W_2(1)$ & $W_1(2)+W_2(1)$ & $\emptyset$ & $W_2(1)$ & $p_{11}$ & 3 \\
        \hline
        12& $W_1(1)+W_2(2)$ & $W_1(2)+W_2(2)$ & $\emptyset$ & $W_2(2)$ & $p_{12}$ & 3 \\
        \hline
        \end{tabular}
    \end{subtable}\\

\vspace{10pt}

\begin{subtable}[t]{\textwidth}
    \centering
    \caption{Message length $L=1$}
    \begin{tabular}{ |c|c|c|c|c|c|c| }
    \hline
    Option & Database $1$ & Database $2$ & Database $3$ & Database $4$ & Probability & \thead{Download \\cost} \\
    \hline
    1 & $W_1(1)$ & $\emptyset$ & $\emptyset$ & $\emptyset$ & $p_1$ & 1 \\
    \hline
    2 & $\emptyset$ & $W_1(1)$ & $\emptyset$ & $\emptyset$ & $p_2$ & 1 \\
    \hline
    3 & $\emptyset$ & $\emptyset$ & $W_1(1)$ & $\emptyset$ & $p_3$ & 1 \\
    \hline
    4 & $\emptyset$ & $\emptyset$ & $\emptyset$ & $W_1(1)$ & $p_4$ & 1 \\
    \hline
    5 & $W_2(1)$ & $W_1(1)+W_2(1)$ & $\emptyset$ & $\emptyset$ & $p_5$    & 2 \\
    \hline
    6 & $\emptyset$ & $W_2(1)$ & $W_1(1)+W_2(1)$ & $\emptyset$ & $p_6$    & 2 \\
    \hline
    7 & $\emptyset$ & $\emptyset$ & $W_2(1)$ & $W_1(1)+W_2(1)$ & $p_7$    & 2 \\
    \hline
    8 & $W_1(1)+W_2(1)$ & $\emptyset$ & $\emptyset$ & $W_2(1)$ & $p_8$    & 2 \\
    \hline
    \end{tabular}
\end{subtable}
\end{table*}

From Lemma \ref{lemma2}, we have the following theorem for the alternative PIR scheme. We denote the information leakage in the alternative PIR scheme by $\rho_{\alpha}^{alt}$ when R\'{e}nyi divergence of order $\alpha$ is used.
\begin{theorem}\label{Theorem2}
    The optimal tradeoff between the information leakage measured by the R\'{e}nyi divergence $\rho_{\alpha}^{alt}$ of order $\alpha$, where  $0<\alpha<\infty$ and $\alpha\ne1$ and the expected normalized download cost $D$, where $1\leq D\leq \frac{1}{C}$ on the proposed alternative probabilistic PIR scheme is given as
    \begin{equation}
    \begin{aligned} \label{thm2}
        \rho_{\alpha}^{alt} = & ~\frac{1}{\alpha-1}\log
        \biggr[ N \left \{ \frac{1-L(D-1)}{N} \right \}^{\alpha} \\
        & + (N(L+1)^{K-1}-N)\left \{\frac{L(D-1)}{N(L+1)^{K-1}-N}\right \}^{\alpha} \biggr] \\
        & + \log N(L+1)^{K-1}.
    \end{aligned}
    \end{equation}
    Then the information leakage-download cost pairs establish the optimal tradeoff. In the special cases of $\alpha=1$ and $\alpha\to\infty$, $\rho_1^{alt}$ and $\rho_{\infty}^{alt}$ are given as
    \begin{equation}
    \begin{aligned} \label{thm2_KL}
        \rho_{1}^{alt} = ~& \left\{1-L(D-1)\right\}\log \frac{1-L(D-1)}{N} \\                          &+ L(D-1)\log \frac{L(D-1)}{N(L+1)^{K-1}-N} \\
                           &+ \log N(L+1)^{K-1},
    \end{aligned}
    \end{equation}
    and
    \begin{equation}
    \begin{aligned} \label{thm2_infty}
        \rho_{\infty}^{alt} =& \log\frac{1-L(D-1)}{N} +\log N(L+1)^{K-1},
    \end{aligned}
    \end{equation}
    respectively.
\end{theorem}

\begin{IEEEproof}
Proof is given by substituting the optimal solution from Lemma \ref{lemma2} in the objective functions of \eqref{min_alt}, \eqref{min_alt_KL}, and \eqref{min_alt_infty}.
\end{IEEEproof}

\subsection{Numerical Analysis}
\begin{figure*}[t]
    \centering
    \begin{subfigure}[t]{0.48\textwidth}
        \centering
        \includegraphics[width=\textwidth]{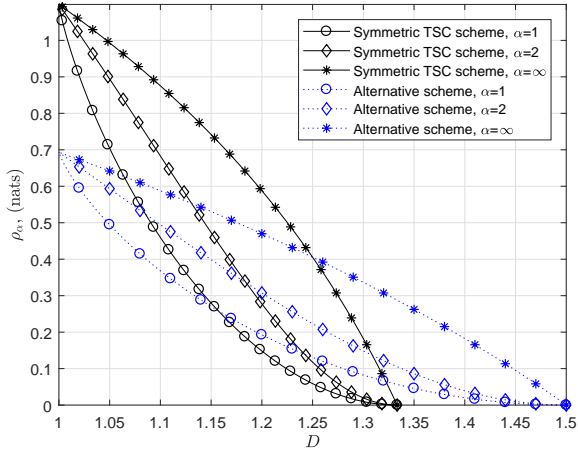}
        \caption{Without normalization of information leakage}
    \end{subfigure}
    \begin{subfigure}[t]{0.48\textwidth}
        \centering
        \includegraphics[width=\textwidth]{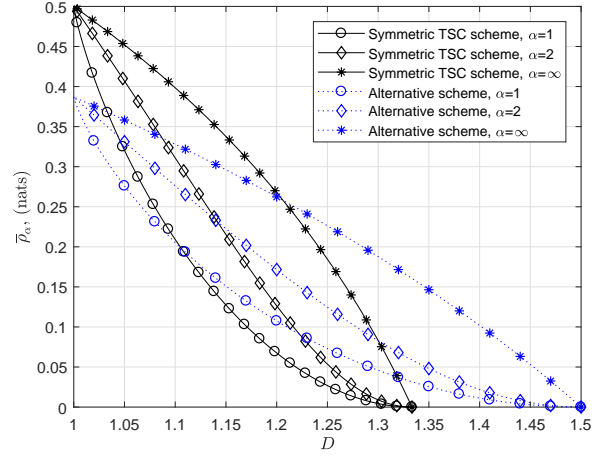}
        \caption{With normalization of information leakage}
    \end{subfigure}
\caption{Optimal tradeoffs between information leakage and download cost for the two PIR schemes with $N=3,K=2$.}    
\label{graphN3K2_alt_nor}
\end{figure*}

By using Theorems \ref{Theorem1} and \ref{Theorem2}, we compare the optimal tradeoffs achieved from two probabilistic PIR schemes with the same parameters $N=3, K=2$: the symmetric TSC scheme in Table \ref{table_N3K2} and the alternative PIR scheme in Table \ref{table_N3K2_alt}. Fig. \ref{graphN3K2_alt_nor}(a) shows the optimal tradeoff between the information leakage and the download cost. Note that the three typical orders of $\alpha=1,2$, and $\infty$ are presented.

\begin{figure*}[t]
    \centering
    \begin{subfigure}[t]{0.48\textwidth}
        \centering
        \includegraphics[width=\textwidth]{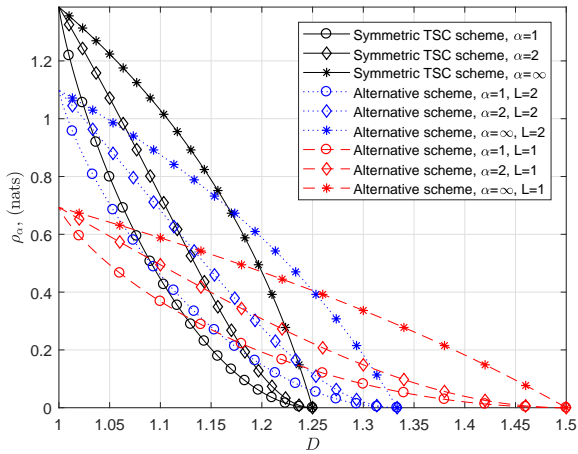}
        \caption{Without normalization of information leakage}
    \end{subfigure}
    \begin{subfigure}[t]{0.48\textwidth}
        \centering
        \includegraphics[width=\textwidth]{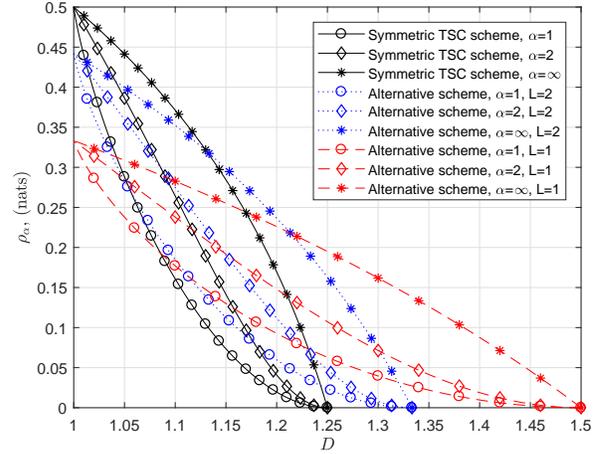}
        \caption{With normalization of information leakage}
    \end{subfigure}
\caption{Optimal tradeoffs between information leakage and download cost for the two PIR schemes with $N=4,K=2$.}    
\label{graphN4K2_alt_nor}
\end{figure*}

We observe that the information leakage of the proposed alternative PIR scheme can be lower than the information leakage of the TSC scheme, i.e., $\rho_{\alpha}^{alt} \le \rho_{\alpha}$ for the lower download cost region. In other words, the alternative PIR scheme achieves a download cost that is smaller than that of the conventional symmetric TSC scheme when we assume high allowance for information leakage. Therefore, when download cost is prioritized over information leakage, the alternative PIR scheme is preferred. However, since the two PIR schemes have the different option sizes of $N^K$ and $N(L+1)^{K-1}$, respectively, a normalized version of information leakage can be considered for a more fair comparison. The normalization of \eqref{D_KL_PU} as an information leakage measure is defined as
\begin{align}\label{D_KL_PU_norm}
    \bar{\rho}_{\alpha} &=\frac{D_{\alpha}(P\parallel U)}{H_\alpha(U)}.
\end{align}
Fig. \ref{graphN3K2_alt_nor}(b) shows the same example of $N=3, K=2$ with normalized information leakage. Note that there still exists the low download cost region satisfying $\bar{\rho}_{\alpha}^{alt} < \bar{\rho}_{\alpha}$. 

We present another example of $N=4, K=2$. Compared to the previous Table \ref{table_N4K2} of the symmetric TSC scheme, query structures of the alternative PIR scheme are shown in Tables \ref{table_N4K2_alt}(a) and (b). Since $1\leq L < N-1$, there are two possible message lengths, $L=1$ and $L=2$. We present the optimal tradeoffs in Fig. \ref{graphN4K2_alt_nor}(a) for both message lengths along with the symmetric TSC scheme, and their normalized information leakage versions are shown in Fig. \ref{graphN4K2_alt_nor}(b) as well. Note that as the message length $L$ shortens, the alternative PIR scheme performs better in the lower download cost region. In a practical scenario, therefore, a user may want to choose among $N-1$ possible schemes, including the symmetric TSC scheme according to the target download cost.

\begin{figure}[t]
\centering
    \includegraphics[scale=0.58]{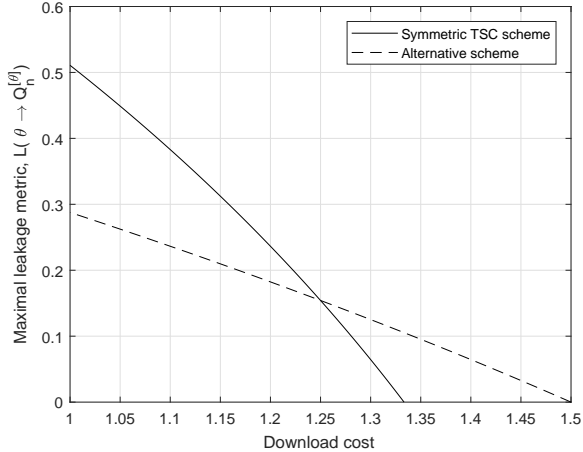}
\caption{Optimal tradeoffs between maximal leakage metric and download cost for the TSC scheme and the alternative PIR scheme with $N=3,K=2$.}    
\label{graphN3K2_alt_maxL}
\end{figure}

Finally, we investigate the information leakage of the alternative PIR scheme measured by the maximal leakage metric adopted in \cite{WPIR_maxL}. Fig. \ref{graphN3K2_alt_maxL} shows the same characteristics as those earlier tradeoffs. When the download cost needs to be low, the alternative PIR scheme incurs lower maximal leakage than that of the symmetric TSC scheme.

\section{Conclusion}
In this paper, we introduced R\'{e}nyi divergence as a measure of information leakage in PIR schemes with probabilistic query structures. We then formulated convex optimization problems to characterize the optimal tradeoff between information leakage and the download cost.
Interestingly, the probability distribution minimizing the Renyi divergence is identical to the distribution minimizing maximal leakage.
We also considered an alternative PIR scheme with shorter message sizes than that of the original TSC scheme. 
The alternative PIR scheme can achieve lower information leakage than that of the TSC scheme for the region of lower download costs.

\appendices
\section{Proof of Lemma \ref{lemma1}: $0<\alpha<\infty$} \label{app.A}
\subsection*{i) For the case of $\alpha>1$;}
The problem in \eqref{minTSC} can be reduced to
\begin{equation} \label{minTSC_alpha>1}
\begin{aligned}
    \underset{P}{\text{minimize}}~~~~~ &\sum_{m=1}^{N^K}{{p_m^{\alpha}}} \\
    \text{subject to}~~~~~ &\frac{1}{N-1}\sum_{m=1}^{N^K}{p_m d_m} = D, \\
                    &\sum_{m=1}^{N^K}{p_m} = 1.
\end{aligned}
\end{equation}
The Lagrangian for problem \eqref{minTSC_alpha>1} is given as
\begin{equation}
\begin{aligned}
    \mathfrak{L}(P,\lambda,\nu) =& \sum_{m=1}^{N^K}{{p_m^{\alpha}}} + \lambda\left(\frac{1}{N-1}\sum_{m=1}^{N^K}{p_m d_m}-D\right) \\ & +\nu\left(\sum_{m=1}^{N^K}{p_m}-1\right),
\end{aligned}
\end{equation}
where $\lambda$ and $\nu$ are the Lagrange multipliers. Taking partial derivatives, we obtain
\begin{equation}\label{der_A}
\begin{aligned}
    \frac{\partial \mathfrak{L}}{\partial p_m} &= \alpha p_m^{\alpha-1} + \frac{1}{N-1}\lambda d_m + \nu, ~m\in [1:N^K].
\end{aligned}
\end{equation}
Equating derivatives in \eqref{der_A} with zero gives the solution to $P$ as
\begin{align}
    p_m = \left\{-\frac{1}{\alpha}\left( \frac{1}{N-1}\lambda d_m + \nu \right )\right\}^{\frac{1}{\alpha-1}}, \\ ~m\in [1:N^K], \label{prob_alpha>1}
\end{align}
where $P=(p_1,\cdots,p_{N^K})$ satisfies
\begin{equation} \label{const}
\begin{aligned}
    \frac{1}{N-1}\sum_{m=1}^{N^K}{p_m d_m} &= D, \\
    \sum_{m=1}^{N^K}{p_m} &= 1.
\end{aligned}
\end{equation}
For a given set of download cost $d_m, m\in[1:N^K]$ and the target download cost $D$, we can solve \eqref{const} to find $\lambda$ and $\nu$. After then, the optimal probability allocation \eqref{prob_alpha>1} can be obtained.

In general, it is hard to find a closed-form solution for $P$ in the nonlinear equation. This is because there exists no algebraic solution to general quintic or higher order equations with arbitrary polynomial coefficients. Specifically, if $N^K\ge5$ with arbitrary $d_1,\dots,d_{N^K}$, we cannot solve \eqref{const} for $\lambda$ and $\nu$, in a closed-form. Therefore, a numerical method should be implemented.
However, since the probabilistic query structure of the symmetric TSC scheme has only two types of download costs, it is possible to solve the optimization problem explicitly. In the symmetric TSC scheme, we have
\begin{align}
	d_m =
        \begin{cases}
            N-1, & m\in[1:N], \\
            N,   & m\in [N+1:N^K],
        \end{cases} \label{d_m}
\end{align}
then \eqref{prob_alpha>1} becomes
\begin{align}
	p_m =
        \begin{cases}
            \left\{-\frac{1}{\alpha}\left(\lambda+\nu\right)\right\}^\frac{1}{\alpha-1}, & m\in[1:N], \\
            \left\{-\frac{1}{\alpha}\left(\frac{N}{N-1}\lambda+\nu\right)\right\}^\frac{1}{\alpha-1},   & m\in [N+1:N^K].
        \end{cases}
\end{align}
Now let $\tau_1=p_m$ for $m\in[1:N]$ and $\tau_2=p_m$ for $m\in[N+1:N^K]$. Then \eqref{const} is written as
\begin{equation}
\begin{aligned}
        \frac{1}{N-1}\{N(N-1)\tau_1 +(N^K-N)N\tau_2\} &= D, \\
        N\tau_1+(N^K-N)\tau_2 &= 1,
\end{aligned}
\end{equation}
from which the solutions for $\tau_1$ and $\tau_2$ are obtained as
\begin{equation}
\begin{aligned} \label{taus}
    \tau_1 &= \frac{1-(N-1)(D-1)}{N}, \\
    \tau_2 &= \frac{(N-1)(D-1)}{N^K-N}.
\end{aligned}
\end{equation}

\subsection*{ii) For the case of $0<\alpha<1$;}
The problem in \eqref{minTSC} can be reduced to
\begin{equation} \label{minTSC_alpha<1}
\begin{aligned}
    \underset{P}{\text{minimize}}~~~~~ &-\sum_{m=1}^{N^K}{{p_m^{\alpha}}} \\
    \text{subject to}~~~~~ &\frac{1}{N-1}\sum_{m=1}^{N^K}{p_m d_m} = D, \\
                    &\sum_{m=1}^{N^K}{p_m} = 1,
\end{aligned}
\end{equation}
where its Lagrangian is given as
\begin{equation} \label{Lag_B}
\begin{aligned}
    \mathfrak{L}(P,\lambda,\nu) =& -\sum_{m=1}^{N^K}{{p_m^{\alpha}}} + \lambda\left(\frac{1}{N-1}\sum_{m=1}^{N^K}{p_m d_m}-D\right) \\ & +\nu\left(\sum_{m=1}^{N^K}{p_m}-1\right).
\end{aligned}
\end{equation}
Similar to \eqref{prob_alpha>1}, we have
\begin{align}
    p_m = \left\{\frac{1}{\alpha}\left( \frac{1}{N-1}\lambda d_m + \nu \right )\right\}^{\frac{1}{\alpha-1}}, ~m\in [1:N^K], \label{prob_alpha<1}
\end{align}
where $P=(p_1,\cdots,p_{N^K})$ satisfies \eqref{const}. Solving for $p_m$ is identical to the case of $\alpha>1$ with two types of probabilities
\begin{align}
	p_m =
        \begin{cases}
            \left\{\frac{1}{\alpha}\left(\lambda+\nu\right)\right\}^\frac{1}{\alpha-1}, & m\in[1:N], \\
            \left\{\frac{1}{\alpha}\left(\frac{N}{N-1}\lambda+\nu\right)\right\}^\frac{1}{\alpha-1},   & m\in [N+1:N^K],
        \end{cases}
\end{align}
and yields the same solution as in \eqref{taus}.

\subsection*{iii) For the case of $\alpha=1$;}
The problem in \eqref{minTSC_KL} is reduced to
\begin{equation}
\begin{aligned}
    \underset{P}{\text{minimize}}~~~~~ &\sum_{m=1}^{N^K}{p_m \log {p_m}} \\
    \text{subject to}~~~~~ &\frac{1}{N-1}\sum_{m=1}^{N^K}{p_m d_m} = D, \\
                    &\sum_{m=1}^{N^K}{p_m} = 1,
\end{aligned}
\end{equation}
where its Lagrangian is given as
\begin{equation}
\begin{aligned}
    \mathfrak{L}(P, & \lambda,\nu) \label{Lag_C} =  \sum_{m=1}^{N^K}{p_m \log p_m}  \\
    &+ \lambda\left(\frac{1}{N-1}\sum_{m=1}^{N^K}{p_m d_m}-D\right) +\nu\left(\sum_{m=1}^{N^K}{p_m}-1\right).
\end{aligned}
\end{equation}
Taking partial derivatives, we obtain
\begin{equation} \label{der_C}
\begin{aligned}
    \frac{\partial \mathfrak{L}}{\partial p_m} &= \log p_m + 1 + \frac{1}{N-1}\lambda d_m + \nu, ~m\in [1:N^K], \\
\end{aligned}
\end{equation}
and equating them with zero gives the solution to $P$ as
\begin{align}
    p_m = \frac{1}{e^{\frac{1}{N-1}\lambda d_m+ \nu + 1}}, ~m\in [1:N^K], \label{prob_alpha=1}
\end{align}
where $P=(p_1,\cdots,p_{N^K})$ satisfies \eqref{const}. Similar to the case of $\alpha>1$ and $0<\alpha<1$, we have two types of probabilities as
\begin{align}
	p_m =
        \begin{cases}
            \frac{1}{e^{\lambda+\nu+1}}, & m\in[1:N], \\
            \frac{1}{e^{\frac{N}{N-1}\lambda+\nu+1}}. & m\in [N+1:N^K].
        \end{cases}
\end{align}
Therefore, with the constraints \eqref{const}, the same solution as in \eqref{taus} is obtained.



\section{Proof of Lemma \ref{lemma1}: $\alpha\to\infty$} \label{app.B}
The problem in \eqref{minTSC_infty} is rewritten as 
\begin{equation}
\begin{aligned}
    \underset{P,~t}{\text{minimize}}~~~~~   &t \\
    \text{subject to}~~~~~ 
        &p_m\leq t, ~m\in [1:N^K], \\
        &\frac{1}{N-1}\sum_{m=1}^{N^K}{p_m d_m} = D, \\
        &\sum_{m=1}^{N^K}{p_m} = 1.
\end{aligned}
\end{equation}
The Lagrangian is given as
\begin{equation} \label{Lag_infty}
\begin{aligned}
    \mathfrak{L}(P, & t,\eta,\lambda,\nu) = 
    t + \sum_{m=1}^{N^K}{\eta_m(p_m-t)} \\
    &+ \lambda\left(\frac{1}{N-1}\sum_{m=1}^{N^K}{p_m d_m}-D\right) +\nu\left(\sum_{m=1}^{N^K}{p_m}-1\right),
\end{aligned}
\end{equation}
where $\eta_m$, $\lambda$, and $\nu$ are the Lagrange multipliers. Taking partial derivatives and equating them with zero, we obtain
\begin{align}  
    \label{der_infty1} \frac{\partial \mathfrak{L}}{\partial p_m} &= \eta_m + \frac{1}{N-1}\lambda d_m + \nu =0, m\in [1:N^K],~~ \\
    \label{der_infty2} \frac{\partial \mathfrak{L}}{\partial t} &= 1-\sum_{m=1}^{N^K}{\eta_m} =0.
\end{align}
Since we have only two types of $d_m$ as in \eqref{d_m}, $\eta_m$ in \eqref{der_infty1} takes only two values as
\begin{equation} \label{nu}
\begin{aligned}
    \eta_m =
    \begin{cases}
        -\lambda-\nu, &m\in[1:N],\\
        -\frac{N}{N-1}\lambda-\nu, &m\in[N+1:N^K].\\
    \end{cases}
\end{aligned}    
\end{equation}
From the complementary slackness, we have
\begin{equation} \label{comp_slack}
\begin{aligned}
    \eta_m (p_m-t) = 0,    
\end{aligned}
\end{equation}
for $\eta_m \ge 0$ by definition. 

If $\eta_m>0$ for all $m\in[1:N^K]$, then $P$ is uniform with $p_m=t$. 
Otherwise, if we let $\varOmega\subset[1:N^K]$ such that $\eta_m>0$ for $m\in\varOmega$, then $p_m=t$ for $m\in\varOmega$.
Since $\eta_m$ is in two values as in \eqref{nu}, we have $\eta_m=0$ and $p_m\leq t$ for $m\in\varOmega^C$. In other words, $\varOmega$ and $\varOmega^C$ are denoted as
\begin{equation}
\begin{aligned}
    \varOmega   &= \{m:\eta_m>0,p_m=t,     m\in[1:N^K]\},\\
    \varOmega^C &= \{m:\eta_m=0,p_m\leq t, m\in[1:N^K]\} \\
                &=[1:N^K]\backslash\varOmega,
\end{aligned}
\end{equation}
respectively.
Therefore, we can think of the two cases on $\varOmega$: $\varOmega=[1:N]$ or $\varOmega=[N+1:N^K]$. By intuition, we first choose the former case of $\varOmega=[1:N]$ whose download costs are lower and allocate the highest $p_m=t$ for $m\in[1:N]$. From the constraints in the primal problem and \eqref{d_m},
\begin{equation}
\begin{aligned}
    Nt + \frac{N}{N-1} \sum_{m=N+1}^{N^K}{p_m} &= D, \\
    Nt + \sum_{m=N+1}^{N^K}{p_m} &= 1,
\end{aligned}
\end{equation}
from which we can solve for
\begin{equation}
\begin{aligned}
    t &=\frac{1-(N-1)(D-1)}{N},
\end{aligned}
\end{equation}    
and
\begin{equation} \label{p_m_low}
\begin{aligned}    
    \sum_{m=N+1}^{N^K}{p_m} &=(N-1)(D-1).
\end{aligned}
\end{equation}
Therefore, we have
\begin{equation}
\begin{aligned}
    p_m=\frac{1-(N-1)(D-1)}{N}, ~m\in[1:N],
\end{aligned}
\end{equation}
and $p_m$ for $m\in [N+1:N^K]$ can take any distribution satisfying \eqref{p_m_low} and $p_m\leq \frac{1-(N-1)(D-1)}{N}$.

On the other hand, if we choose $\varOmega=[N+1:N^K]$, we can solve for
\begin{equation}
\begin{aligned}
    p_m = \frac{(N-1)(D-1)}{N^K-N},~m\in[N+1:N^K],
\end{aligned}
\end{equation}    
and $p_m$ for $m\in [1:N]$ can take any distribution satisfying $\sum_{m=1}^{N}{p_m} =1-(N-1)(D-1)$
and $p_m\leq\frac{(N-1)(D-1)}{N^K-N}$.
In this case, we cannot find such distribution in the operational range of $D\in [1,\frac{1}{C})$. 
This is easily verified by letting $D=1$, which leads to
\begin{equation}
\begin{aligned}   
    p_m =0,&~m\in[N+1:N^K],
\end{aligned}
\end{equation}
and $p_m\leq0$ for $m\in[1:N]$ cannot satisfy $\sum_{m=1}^{N}{p_m} =1$. Therefore, $\varOmega=[N+1:N^K]$ is not for the solution.


\ifCLASSOPTIONcaptionsoff
  \newpage
\fi

\end{document}